\documentclass[fleqn,usenatbib]{mnras}

\usepackage{newtxtext,newtxmath}

\usepackage[T1]{fontenc}

\DeclareRobustCommand{\VAN}[3]{#2}
\let\VANthebibliography\thebibliography
\def\thebibliography{\DeclareRobustCommand{\VAN}[3]{##3}\VANthebibliography}

\usepackage{graphicx}	
\usepackage{amsmath}	
\def\rx{RX J0123.4-7321}
\def\m2002 {[M2002] SMC 81035}
\newcommand{\bexrb}{BeXRB}
\newcommand{\bexrbs}{BeXRBs}

\title[RX J0123.4-7321]{RX J0123.4-7321 - the story continues: major circumstellar disk loss and recovery.}

\author[M. J. Coe et al.]{M.~J. Coe$^{1}$\thanks{E-mail: mjcoe@soton.ac.uk},
 A. Udalski$^{2}$, J.~A. Kennea$^{3}$, P.~A. Evans$^{4}$\\
$^{1}$Physics \& Astronomy, The University of Southampton, SO17 1BJ, UK\\
$^{2}$Astronomical Observatory, University of Warsaw, Al. Ujazdowskie 4, 00-478 Warszawa, Poland\\
$^{3}$Department of Astronomy and Astrophysics, The Pennsylvania State University, 525 Davey Lab, University Park, PA 16802, USA\\
$^{4}$University of Leicester, Astrophysics Group, School of Physics \& Astronomy, University Road, Leicester LE1 7RH, UK\\
}

\date{Accepted XXX. Received YYY; in original form ZZZ}

\pubyear{2021}

\begin{document}
\label{firstpage}
\pagerange{\pageref{firstpage}--\pageref{lastpage}}
\maketitle

\begin{abstract}
\rx ~is a well-established Be star X-ray binary system (\bexrb) in the Small Magellanic Cloud (SMC). Like many such systems the variable X-ray emission is driven by the underlying behaviour of the mass donor Be star. Previous work has shown that the optical and X-ray were characterised by regular outbursts at the proposed binary period of 119 d. However around February 2008 the optical behaviour changed substantially, with the previously regular optical outbursts ending. Reported here are new optical (OGLE) and X-ray (Swift) observations covering the period after 2008 which suggest an almost total circumstellar disc loss followed by a gradual recovery. This indicates the probable transition of a Be star to a B star, and back again.  However, at the time of the most recent OGLE data (March 2020) the characteristic periodic outbursts had yet to return to their early state, indicating that the disk still had some re-building yet to complete.
\end{abstract}




\section{Introduction}

\bexrb\ are a large sub-group of the well-established category of High Mass X-ray Binaries (HMXB) characterised by being a binary system consisting of a massive mass donor star, normally an OBe type, and an accreting compact object, a neutron star. They are particularly prevalent in the Small Magellanic Cloud (SMC) which contains the largest known collection of \bexrbs\ . ~Whilst catalogues have been produced listing such systems in the SMC (for example \cite{ck2015, hs2016}), it clear that the complex interactions between the two stars continues to produce unexpected surprises. In particular, the behaviour of the mass donor B-type star is major driver in the observed characteristics of such systems.

The source that is the subject of this paper, \rx, was identified early on as an X-ray source in the SMC \citep{haberl2000}, but it took over another decade before its identity as \bexrb\ was clearly established by \cite{sturm2013}. In their paper they not only identified the X-ray source as a partner to the Be star \m2002 , but also used optical photometry from the OGLE III project to show that the binary period of the system was 119 d. 

Reported here are the successive 10 years of OGLE IV data and 4 years of X-ray observations from the Swift X-ray Telescope (XRT; \citealt{burrows05}) as part of the S-CUBED project \citep{kennea2018}. The new OGLE data show a major change in the behaviour of the Be star as it appears to almost completely lose its circumstellar disk. This represents a rarely seen transition of \m2002 ~from a classic Be star to a regular B-type star, and back again. All within a decade. The X-ray data, which are concurrent with the last few years of OGLE IV, reveal 15 detections of \rx, with the detections extremely well correlated with the phase of the optical outbursts.  These epochs are believed to correspond to times of periastron passage of the neutron star, and the profiles of the optical \& X-ray outbursts are very similar, suggesting a common mechanism related to the periodic distortion of the circumstellar disk.
 
\section{Observations}

\subsection{OGLE}

The OGLE project \citep{Udalski2015}  provides long term I-band photometry with a cadence of 1-3 days. The star [M2002] SMC81035 was observed continuously for nearly 2 decades until COVID-19 restrictions prevented any further observations from March 2020. It is identified in the OGLE catalogue as:\\
\\
OGLE III (I band): smc121.2.50 \\
OGLE IV (I band): smc733.26.24 \\
OGLE III (V band) : smc121.2.v.18 \\
OGLE IV (V band):  smc733.26.v.13 \\

\begin{figure}

	\includegraphics[width=8cm,angle=-00]{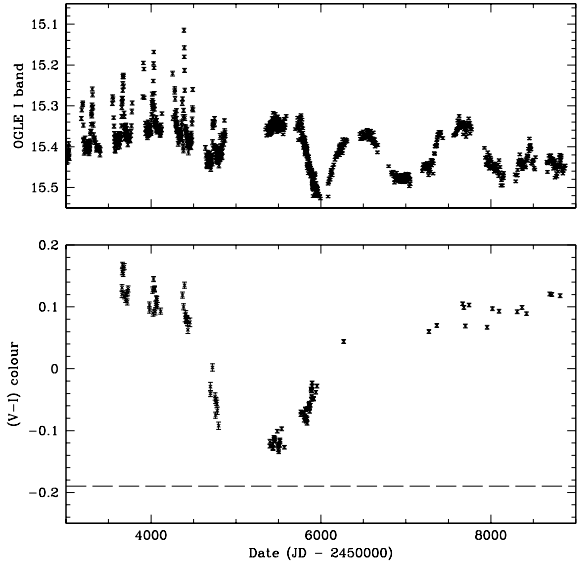}
    \caption{Upper panel : OGLE III and IV data - the transition between the two observing phases occurred around TJD 5000. Lower panel: OGLE (V-I) colour measurements. The dashed horizontal line indicates the predicted observed colour of a B0.5 III star in the SMC.}
    \label{fig:2panels}
\end{figure}

The I band data are shown in their entirety in the top panel of Fig \ref{fig:2panels}. The OGLE III data were previously presented in \cite{sturm2013} who reported overwhelming evidence for the binary period of this system based upon the sharp regular outburst features seen every 119 d. However this striking feature ended around TJD 5000 just as the OGLE III project phase also came to an end. Presented here are the OGLE IV data which add a further 10 years to the story, and indicate a very dramatic change in behaviour pattern from OGLE III. For most of this last decade the regular outbursts are missing completely and only just start to re-emerge around TJD 8500, though not as strongly as before. Nonetheless, the presence of these features ten years later permit a small refinement to be made to the binary period ephemeris that was originally presented in \cite{sturm2013}. The updated ephemeris for the time of the optical outbursts, $T_{opt}$, is:

\begin{equation}
T_{opt} = 54387.1 + N(119.59) ~\textrm{MJD}\label{eq:1}
\end{equation}

The optical counterpart \m2002 ~ is reported to be a B0.7 IIIe star by \cite{massey2002} and \cite{rama2019}. The exact intrinsic colours of such a spectral type are not readily tabulated, \textcolor{black}{but the (V-I) colour of an almost identical B0.5 V star is given in \cite{pm2013} as -0.338.} They do not quote the colours of a giant of the same type, but this difference is given in \citep{weg94} as 0.08. Thus it is expected that the intrinsic colours of a B0.5III will be (V-I) = -0.258. 

The OGLE dust maps of the SMC \citep{skowron2021} enable the precise reddening correction to be made for such an object in the SMC and that is E(V-I)=0.067. Thus the predicted observed colours of \m2002 ~if it were a B-type star B0.5II with no circumstellar disk will be (V-I) = -0.19. The difference between a B0.5 and B1 star is less than 0.01 \citep{pm2013} so any difference between a B0.5 and B0.7 will be negligible for the purpose of this discussion. Thus the calculated colour is indicated in the lower panel of Fig \ref{fig:2panels} as the horizontal dashed line. From this figure it is immediately clear that all the observed OGLE colours for this object lie significantly above the baseline, strongly suggesting the presence of a circumstellar disk adding further reddening to the observed colours.

Clearly if the red excess in the colour of \m2002 ~is interpreted as the existence of a circumstellar disk, then the huge variations in the (V-I) colour seen in Fig \ref{fig:2panels} must also be indicating significant changes in the disk size over the last 20 years. From a substantive disk at the start of this period, through to almost nothing around TJD 5000. Then the star went through a series of re-growth phases to reclaim some of its original status as a probable Be type. This is further discussed below.

\subsection{S-CUBED}

\begin{figure*}
	\includegraphics[width=16cm,angle=-00]{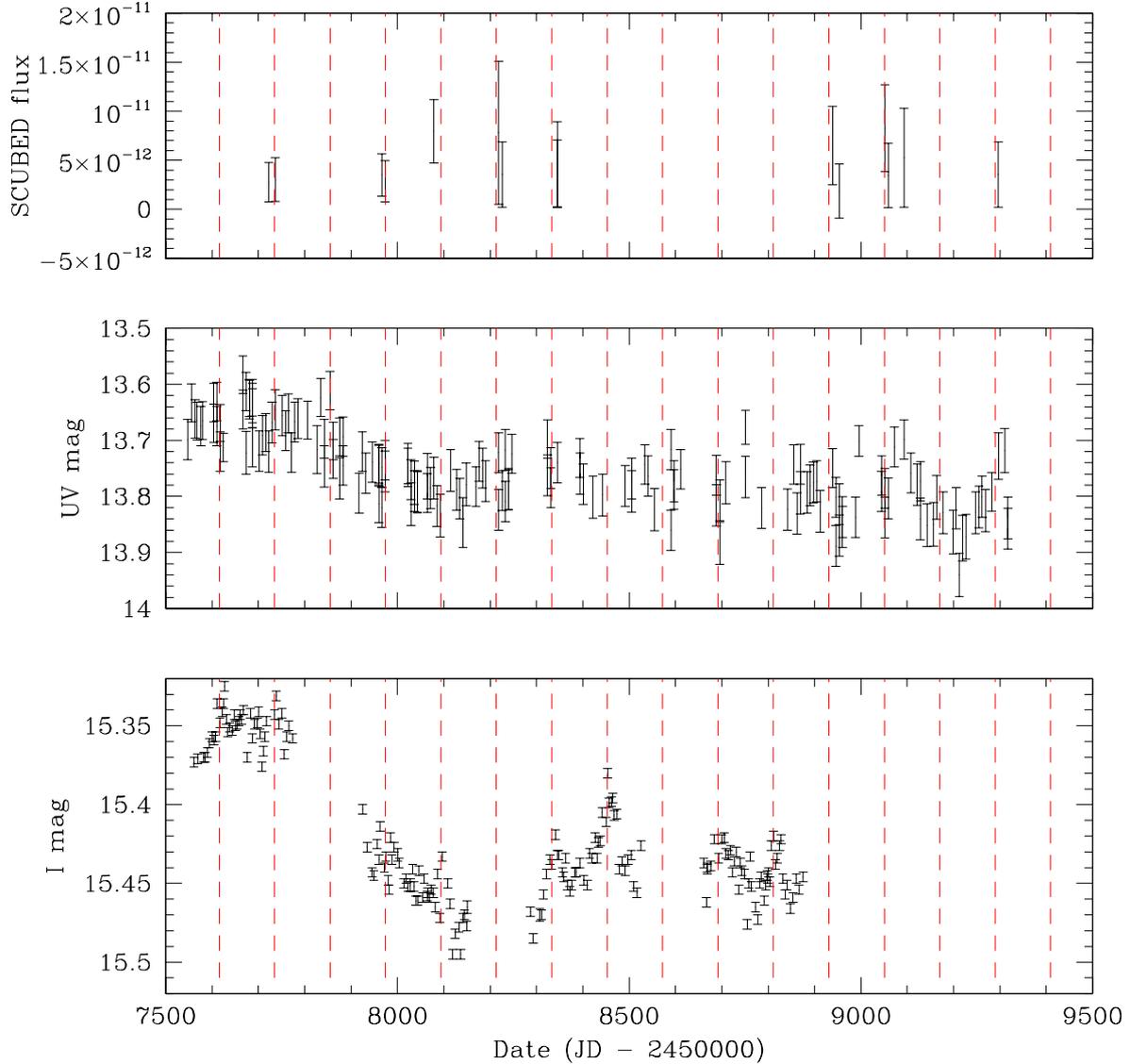}
    \caption{Comparison of the X-ray, UV and optical detections. Upper panel S-CUBED X-ray detections of \rx ~in flux units of ${\rm erg}\cdot {\rm cm}^{-2}\cdot {\rm s}^{-1}$. Centre panel : UVOT magnitudes in uvw1 filter of \m2002 . ~Bottom panel : OGLE IV I-band data of \m2002 . The vertical red dashed lines indicate the dates of expected optical outburst given by Equation \ref{eq:1}.}
    \label{fig:xo}
\end{figure*}

\rx{} was detected by the S-CUBED survey \citep{kennea2018}, a shallow weekly X-ray survey of the optical extent of the SMC by the Swift X-ray Telescope (XRT; \citealt{burrows05}). Individual exposures in the S-CUBED survey are typically 60s long, and occur weekly, although interruptions can occur due to scheduling constraints. Starting with the S-CUBED observation taken on MJD 57722 (29 Nov 2016), S-CUBED detected \rx on several subsequent occasions over the following 4-5 years and internally numbered it SC379. The occasions on which it was detected are listed in Table \ref{tab:x} and shown in the top panel of Fig \ref{fig:xo}, together with the flux measured on each occasion. \textcolor{black}{Non-detections have upper limits in the range $(5-10) \times 10^{-12} {\rm erg}\cdot {\rm cm}^{-2}\cdot {\rm s}^{-1}$.}

S-CUBED observations taken with the UV/Optical Telescope (UVOT; \citealt{Roming05}) were also analysed, utilizing the standard \texttt{uvotmaghist} tool. The middle panel of that figure shows the UVOT measurements in the \textit{uvw1} band (2600\AA), it also indicates all the occasions when the source was observed. Broadly speaking the UV tracks the I-band, but not in the same detail. This is partially because the UVOT measurements lack the same precision as the OGLE data, but also because the circumstellar disc radiates predominately in the red end of the spectrum. These UVOT observations seem to confirm that \m2002 ~ has not yet returned to the earlier much brighter state, but has levelled off at some intermediary phase.

\begin{table}
\caption{Table of X-ray detections of \rx ~by S-CUBED over the range 0.3--10 keV}
\label{tab:x}
\begin{tabular}{cccc}
\hline
Date&X-ray flux&X-ray error& Binary phase\\
MJD&\multicolumn{2}{c}{$10^{-12}$~erg per (sq-cm sec) }&\\
\hline
57722.37&	2.76&		1.34&	0.88\\
57736.24&	3.02&		1.49&	0.00\\
57966.38&	3.50&		1.54&	0.93\\
57973.62&	2.85&		1.42&	0.99\\
58078.46&	7.96&		2.54&	0.86\\
58218.41&	7.82&		4.46&	0.03\\
58225.97&	3.54&		2.03&	0.09\\
58344.77&	4.57&		2.66&	0.09\\
58344.78&	3.63&		2.08&	0.09\\
58939.04&	6.54&		2.83&	0.06\\
58953.16&	1.85&		1.35&	0.18\\
59052.03&	8.26&		3.27&	0.01\\
59059.00&	3.45&		2.00&	0.07\\
59093.00&	5.26&		3.07&	0.35\\
59296.41&	3.52&		2.04&	0.05\\
\hline
\end{tabular}
\end{table}

Assuming no spectral changes, one can sum all the detections together to obtain an average flux value of $(5.63 \pm 0.90) \times 10^{-12} {\rm erg}\cdot {\rm cm}^{-2}\cdot {\rm s}^{-1}$. Using a standard SMC distance of 62~kpc \citep{scowcroft2016} and correcting for absorption fixed at the value derived from \cite{Willingale2013}, this corresponds to a 0.5-10~keV luminosity of $(2.5\pm 0.4) \times 10^{36}$~erg$\cdot {\rm s}^{-1}$.

\section{Discussion}

The OGLE V and I band behaviour is shown in  Fig \ref{fig:2panels}. It is obvious that a fundamental change occurred in this emission pattern around TJD 5000. The source changed from clear regular outburst behaviour to one showing long timescale fluctuations. The (V-I) colour record is hugely important in interpreting this change as it shows the bluest colour around the time of this switch in behavioural patterns. It suggests that the Be star lost almost all its circumstellar disk around TJD 5000--5100, and then took 5--6 years to gradually re-build it. It is only around the end of the OGLE IV data that we see the re-commencement of the OGLE III outburst phenomenon, but still only weakly compared to the early years.

\begin{figure}

	\includegraphics[width=8cm,angle=-00]{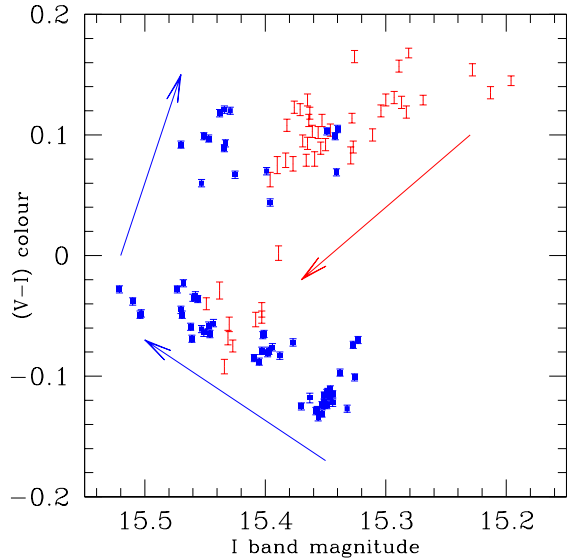}
    \caption{The colour changes observed in the optical companion to \rx ~from both OGLE III and IV data. The red points come from OGLE III and the red arrow indicates the direction of change with time. Blue points are from OGLE IV and the blue arrow indicates the colour evolution with time during that phase.}
    \label{fig:col_evol}
\end{figure}

Fig \ref{fig:col_evol} shows the colour changes occurring between OGLE III and IV as a function of the I band magnitude.In this figure the pathway taken by \m2002 ~over this $\sim$20 year period is indicated. The source begins this time interval in the top right hand corner, at maximum brightness and maximum redness. It then substantially changes over the course of OGLE III heading towards the lower left hand corner - faintest \& bluest emission almost reaching the state of a B0.7 III star with no circumstellar disk. In other words we infer that it has transitioned from a Be star to a normal B-type star. This has occurred on a timescale of just $\sim$5 years.

During OGLE IV we then see the star reverse its position on the colour-magnitude diagram, and in a series of, presumably, mass ejection phases re-builds its circumstellar disk and it resumes its Be type classification. It is interesting to note that the re-building process exhibits itself as large-scale fluctuations on a timescale of $\sim$1000 d. Such a behaviour could be described as a super-orbital modulation, such as that described by \cite{andry2011} and \cite{townsend2020}. In fact the timescales observed here of a binary period of 119 d and a super-orbital period of $\sim$1000 d would fit comfortably on Figure 1 of \cite{townsend2020} in the region occupied by other Be stars in X-ray binary partnerships.

From Fig \ref{fig:2panels} it can be seen that during the time of regular strong outbursts seen in OGLE III the colour (V-I) would change sharply by $\sim$0.07 magnitudes. This changes comes from an increase in the I band of 0.18 magnitudes ($\sim$20\% flux increase), whilst the V band only increased by 0.11 magnitudes ($\sim$10\% flux increase). Assuming an optically thick disk then this is indicative of an increase in the surface area of a disk with an average temperature significantly cooler than that of the parent star, resulting in a redder overall appearance for the system.

\begin{figure}

	\includegraphics[width=8cm,angle=-00]{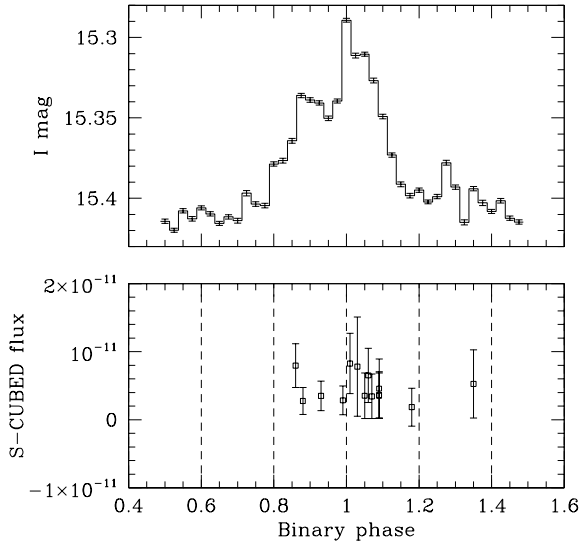}
    \caption{Upper panel : OGLE III data folded at the updated binary ephemeris - see Equation \ref{eq:1}. Lower panel : the X-ray detections of \rx ~as a function of the same binary ephemeris. The X-ray flux is in units of erg~cm$^{-2}$~s$^{-1}$.}
    \label{fig:bin}
\end{figure}

The phases of X-ray detections are shown in Fig \ref{fig:bin} and compared to the OGLE III data folded with the ephemeris given in Equation \ref{eq:1}. It is immediately clear that the X-ray emission is strongly correlated in phase with the OGLE profile, both in width and peak position. The gravitational pull of the arriving neutron star distorts the surface area of the optically thick circumstellar disk increasing the I band emission. At the same time the larger disk is providing material to accrete on to the neutron star triggering the X-ray emission.

To estimate the probable dimensions of the disk and neutron star orbit some assumptions are needed. In the absence of any direct H$\alpha$ equivalent width measurements it is necessary to assume that \m2002 ~is a typical Be star, and we can estimate the circumstellar disk size from its spectral classification, B0.7III. \cite{rivinius2013} quotes the size of the H$\alpha$ emitting disk for a very similar B0.5IVe star as having a typical radius of $7.1R_{*}$. This translates into a disk size of $(3-7) \times 10^{10}$ m for such a star with luminosity class in the range III - V. Also assuming, for simplicity, that the neutron star is in a circular orbit, then the period of 119 d permits an estimate of the orbital radius to be $2 \times10^{11}$ m. That is $\sim$10 times larger than the H$\alpha$ disk size, but as \cite{rivinius2013} points out the actual disk size may well extend beyond the H$\alpha$ emitting region by a significant amount. But, broadly speaking, this is what Smooth Particle Hydrodynamic simulations of such systems predict \citep{okazaki2001,brown2019} - the neutron star orbiting just outside the disk and constraining further disk expansion.

However, the orbital behaviour of the neutron star in this system clearly is not circular, but rather it must exhibit a high degree of eccentricity. Fig \ref{fig:bin} shows that both the optical and X-ray outburst behaviour is restricted to binary phase range 0.8--0.2. Though it is not possible to quantify the degree of eccentricity from those profiles alone, it is obviously very similar to that seen in another system SXP 756 \citep{coe2004}. That system has a much longer binary period of 394 d, but the optical behaviour is remarkably similar to the source in this paper. The optical outbursts are even sharper, extending over a binary phase range of just 0.2. It is therefore highly likely that both of these systems are clear examples of high eccentricity orbits. \cite{townsend2011} demonstrated a significant correlation between the binary period and the eccentricity (where that had been reliably measured as part of an orbital solution) for \bexrb ~systems.  From their Fig 6 it can be predicted that the eccentricity of \rx ~could be in the range 0.4-0.5. Such a high eccentricity would comfortably explain the behaviour seen in this system.

Finally we note from Fig \ref{fig:xo} that though X-ray detections are reported for \textcolor{black}{many} periastron passages of the neutron star, there are some missed occasions. They are missed even though S-CUBED clearly carried out observations at the appropriate times. This is almost certainly just indicative that the X-ray flux levels emitted during these Type I outbursts from \rx ~are close to the detection threshold of the XRT telescope on Swift with just a 60 s exposure. A small reduction in the X-ray strength could easily lead to a missed detection. In fact, Fig \ref{fig:xo} clearly shows evidence for X-ray activity during many binary cycles, even when the I-band flux had diminished around MJD 58000-58400. So, though the circumstellar disk was obviously greatly reduced at these times there was always enough material in the vicinity of the neutron star orbit at periastron to trigger accretion. This is further evidence that the eccentricity of this system must be very high indeed, permitting the neutron star to approach exceptionally close to the Be star each periastron passage.

\section{Conclusions}

Though \rx ~was previously well established as a member of the \bexrb ~family in the SMC, its behaviour has dramatically changed since it was last studied. The new OGLE optical colour data (V-I), combined with the I-band magnitude, strongly indicate a major disk loss episode in this Be star lasting several years. This was followed by a slow recovery over a similar amount of time as the circumstellar disk was gradually rebuilt. Throughout the last few years of this process the X-ray observations reported here were able to detect Type I outbursts around what is believed to be the time of periastron. Despite the almost complete disk loss, these X-ray detections continued over a 4 year period up to the present. They point strongly to a very eccentric orbit bringing the neutron star in close to the Be star and thereby enabling the accretion of material from the limited remains of the circumstellar disk. It will be interesting to see if the characteristic sharp optical outbursts return in the future once the COVID-19 restrictions permit the resumption of OGLE monitoring.

\section*{Acknowledgements}

The OGLE project has received funding from the National Science Centre, Poland, grant MAESTRO 2014/14/A/ST9/00121 to AU. PAE acknowledges UKSA support. JAK acknowledges support from NASA grant NAS5-00136. This work made use of data supplied by the UK Swift Science Data Centre at the University of Leicester.

\section*{Data Availability}

 All X-ray data are freely available from the NASA Swift archive. The optical data in this article will be shared on any reasonable request to Andrzej Udalski of the OGLE project.



\bibliographystyle{mnras}
\bibliography{references} 



\bsp	
\label{lastpage}
\end{document}